\documentclass[showpacs,twocolumn,prb]{revtex4}
\usepackage{graphicx, amsmath, amsthm, amssymb}
\newcommand{\be}{\begin{equation}}
\newcommand{\ee}{\end{equation}}
\newcommand{\bea}{\begin{eqnarray}}
\newcommand{\eea}{\end{eqnarray}}
\newcommand{\bwt}{\begin{widetext}}
\newcommand{\ewt}{\end{widetext}}
\newcommand{\ra}{\rangle}
\newcommand{\la}{\langle}
\newcommand{\up}{\uparrow}
\newcommand{\dn}{\downarrow}

\newcommand{\real}{\mathbb{R}}
\newcommand{\compl}{\mathbb{C}}
\begin{document}
\title{Electron and Spin Transport in the Presence of Complex Absorbing Potential}
\author{Fatih Do\u gan$^1$, Wonkee Kim$^{1,2}$, C.M. Blois$^{1,3}$, and F. Marsiglio$^1$}
\affiliation{ $^1$Department of Physics, University of Alberta,
Edmonton, Alberta, Canada, T6G~2J1 \\
$^2$Texas Center for Superconductivity, University of Houston,
Houston, Texas 77004 \\
$^3$Department of Mathematics, University of Toronto,
Toronto, Ontario, Canada, M5S~2E6}

\begin{abstract}

We examine the impact of a complex absorbing potential on electron
transport, both in the continuum and on a lattice. This requires the
use of non-Hermitian Hamiltonians; the required formalism is briefly outlined.
The lattice formulation allows us to study the interesting problem of
an electron interacting with a stationary spin, and the subsequent time
evolution of the electron and spin properties as the electron is absorbed after
the initial interaction. Remarkably, the properties of the localized spin
are affected 'at-a-distance' by the interaction of the (now entangled) electron
with a complex potential.

\end{abstract}


\date{\today}

\maketitle

\section{introduction}

Typically, the dynamics of a closed system are described by a Hamiltonian which is
Hermitian and accounts for all possible interactions and degrees of freedom.
However, the task of merely writing down the full Hamiltonian is hopeless
in many cases and might also be superfluous from a practical point of view.
Following this line of thought, a reduced or effective Hamiltonian can be
introduced to describe relevant dynamics phenomenologically. As discussed in
Ref. \onlinecite{muga}, for example, the effective Hamiltonian may or may not be Hermitian,
depending on the phenomena under consideration.
For example, a reduction of the full Hamiltonian often used in nuclear physics is
referred to as the 'optical model'. In order to explain experimental results, this model
uses phenomenological potentials such as a complex absorbing potential.

When a complex absorbing potential is introduced into a system with a real
Hamiltonian, the resulting effective Hamiltonian is not Hermitian,
but symmetric and complex in its matrix representation.
The mathematical properties of complex symmetric matrices are a little more
complicated than Hermitian matrices, as detailed in, for example, Ref. \onlinecite{horn}.
There have been many studies, based on various calculational schemes,
of a complex absorbing potential in the continuum limit.
\cite{vibok,bender,midgley,rasmussen,neumair,santra,moiseyev,moiseyev2,riss93,riss96}
The approach followed in this paper  requires the diagonalization of
a complex symmetric Hamiltonian defined on a lattice,
as will be described below.

The textbook example is the one-dimensional scattering problem in which the
absorption is modeled with an imaginary Dirac-$\delta$ potential.
The effective Hamiltonian is
$H=p^{2}/2m - i\beta\delta(x)$ with $\beta>0$.
Then, the scattering states are given by
 $\phi_{k}(x)=\left[e^{ikx}+R\;e^{-ikx}\right]\theta(-x)+
T\;e^{ikx}\theta(x)$, where $\theta$ is the Heaviside step function.
The reflectance and transmittance amplitude $R$ and $T$ are determined by
matching conditions at $x=0$ to be
$R=-\beta/(2k+\beta)$ and $T=2k/(2k+\beta)$. Clearly,  $|R|^{2}+|T|^{2} < 1$.
In fact, the absorption
probability is defined as $|A|^2 \equiv 1-|R|^{2}-|T|^{2}$. A deeper understanding
of this 'absorption' can be attained by studying the dynamics of a sufficiently broad wave packet;
this will be illustrated below.

It is also possible to consider an absorbing potential in a (one-dimensional) lattice, by allowing
the potential $U_i$ at each site $i$ to be complex.
This approach follows that used for impurity potentials on a
lattice as described in Ref. \onlinecite{kim1}. The reflectance and transmittance amplitude
in the case of a complex potential $U=-i\beta\;(\beta > 0)$ at one site is,
respectively,
\be
{\cal R}=-\frac{\beta\;e^{2ik}}{2\sin(k)+\beta}\;,
\label{R_in_lattice}
\ee
and
\be
{\cal T}=\frac{2\sin(k)}{2\sin(k)+\beta}\;.
\label{T_in_lattice}
\ee
with the lattice constant set to be unity. These relations are the basic building
blocks for any more complicated complex potential, to be further examined in this
paper. In particular, spin-flip potentials with the appropriate modifications can
also be described in this way, to understand how spin interactions (along with some
decoherence effects) can modify a spin current. The question of coherence of spin currents
is an important one; in this paper we make a preliminary attempt to model decoherence of
spin and charge currents. We find an interesting property which highlights the 'entanglement'
properties of spins, in particular. These issues are the subject of
this investigation.

This paper is organized as follows:
In Sec.~II we present the formalism of the dynamics of a wave packet
in the presence of complex absorbing potentials in both the continuum limit
and on a lattice.
Sec.~III. illustrates  how a complex potential influences the time-evolution of
a wave packet as a simple example.
In Sec.~IV we study the spin-dependent scattering problem in the presence of a complex
absorbing potential. The time evolution of an expectation value of an operator
is derived when the Hamiltonian is non-Hermitian. We then demonstrate a
highly non-trivial consequence of the complex potential
on spin transfer. This new result shows that quantum entanglement
plays an important role for spin transfer in the presence of a complex
potential. In Sec.~V we conclude with a summary.

\section{formalism}

We begin by considering a single particle in $d$-dimensional space under the influence of a complex
potential.  As usual, the state space is $L^2 (\mathbb{R}^d)$ equipped with the usual inner product
\be
\la f | g \ra := \int_{\real^d} f^* g .
\label{inner product}
\ee
The Hamiltonian, $H:=p^2 /2m + V$ where $V:\mathbb{R}^d \to \mathbb{C}$ achieves non-real values (on
some set of non-zero measure), is easily seen to be non-symmetric (and therefore non-Hermitian).
Computational complications due to a non-Hermitian Hamiltonian are many; some nuances have been
catalogued in Ref. \onlinecite{moiseyev2} and Ref. \onlinecite{santra}, for example.  We follow
Moiseyev  \cite{moiseyev2} and introduce the \emph{c-product}, $( \cdotp | \cdotp ): L^2(\real ^d)
\times  L^2(\real ^d) \to \compl$, defined by
\be
(f|g):=\int_{\real^d} fg .
\label{c-prod}
\ee
(Note that by the Cauchy-Schwarz inequality, $\lvert(f|g)\rvert = \lvert\la f^*|g\ra\rvert\leq\lVert
f^* \rVert \lVert g \rVert = \lVert f \rVert \lVert g \rVert< \infty$ for $f,g \in L^2(\real^d)$.)
We see that the Hamiltonian exhibits a symmetry with respect to the c-product; that is, for all
functions $f$ and $g$ in the domain of H (the Sobolev space $H^2(\real^d)$),
\be
(f|Hg)=(Hf|g).
\ee
This can be easily shown for $f\in H^2(\real^d)$ and $g\in C_c^\infty (\real^d)$, and a density
argument completes the proof.   From this symmetry, it follows that eigenfunctions of $H$
corresponding to distinct eigenvalues are 'orthogonal' with respect to the c-product (rather than the
inner product [\ref{inner product}] as usual).  That is, if $\psi,\phi\in\L^2(\real^d)$, $\lambda,\xi
\in \compl$ with $\lambda \neq \xi$, and $H\psi=\lambda\psi$, $H\phi=\xi\phi$, then
\be
(\psi|\phi)=0.
\ee
For convenience, we now restrict our domain to a $d$-dimensional cube in $\real^d$ and impose
periodic boundary conditions on the edges.  In doing so, we force the spectrum of $H$ to be discrete.
We look for solutions to the initial value problem,
\be
\begin{cases} i\partial_t \Psi_t = H \Psi_t \\
\Psi_0 = \psi \end{cases}.
\label{IVP}
\ee
We suppose that, as in the case where $H$ is Hermitian, the solution is obtained by acting on the
initial wave function $\phi$ with the time-evolution operator $U(t):=e^{-iHt}$.  Then if the
eigenfunctions of $H$ are denoted $\{\phi_j\}_{j=1}^\infty$ with corresponding complex eigenvalues
$E_j$, and the initial wave function is $\psi=\sum a_j \phi_j$, it follows that the solution to Eq.
(\ref{IVP}) is
\be
\Psi_t = \sum_{j=1}^\infty a_j e^{-iE_j t} \phi_j .
\label{time evolution}
\ee
See Appendix A for a sample calculation of wave-packet evolution under the influence of a complex
potential.

In a lattice with $N$ sites, we use the tight-binding Hamiltonian:
\be
H=-t_{0}\sum_{<i,j>\sigma}C^{+}_{i\sigma}C_{j\sigma}+
\sum_{j\in {\cal I},\sigma}U_{j}C^{+}_{j\sigma}C_{j\sigma}\;,
\ee
where $C^{+}_{i\sigma}$ creates an
electron with a spin $\sigma$ at a site $i$,
$t_{0}$ is a hopping amplitude
between the nearest neighbor sites and is set to be unity,
$U_{j}$ is a complex potential at site $j$, and ${\cal I}$ represents
a set of the potential sites. For complex absorbing potentials,
$U_{j}=-i\beta_{j}$, with $\beta_j \geq 0$. Varying $U_{i}$ and ${\cal I}$, one can build
various effective potentials. For example, in order to study
the dynamics of a wave packet and compare the
reflectance [Eq. (\ref{R_in_lattice})] and the transmittance [Eq. (\ref{T_in_lattice})]
given in the Introduction, we choose a single-site potential with $U_j=-i\beta\delta_{0,j}$.

In the lattice, the state space is finite-dimensional; the Hamiltonian can be represented
by a complex symmetric matrix.  Recall that an $N\times N$ matrix with entries in a field ($\compl$)
is diagonalisable
if and only if it has $N$ linearly independent eigenvectors.

An initial electron wave packet with spin up can be written as
\be
|\Psi(0)\ra = \sum_{i=1}^{N}\varphi_{i}C^{+}_{i\up}|0\ra\;,
\ee
where $C^{+}_{i\up}$ acts on the vacuum $|0\ra$
to create an electron with spin up,
and $|\varphi_{i}|^{2}$ is the initial probability to
find such an electron at the
site $i$. Thus in the matrix representation
\be
|\Psi(0)\ra  \doteq \left ( \begin{array}{c}
\varphi_{1}(0)\\
\varphi_{2}(0)\\
\vdots\\
\varphi_{N}(0) \end{array} \right )
\ee
For the purposes of this paper, $|\Psi(0)\ra$ is a Gaussian wave
packet which is far away from (and broad when compared to) the region
of non-zero potential.
In order to find the time evolution of this wave packet, we solve for
all eigenvectors $|n\ra$ and corresponding eigenvalues $E_{n}$ of the
Hamiltonian matrix (and hence verify its diagonalisability).
Now the time-evolution of the wave packet on the lattice
is given by
\be
|\Psi(t)\ra = \sum^{N}_{n=1}(n|\Psi(0)\ra e^{-iE_{n}t}|n\ra\;
\ee
where $(n|$ is the matrix transpose of $|n\ra$.  This can be seen by
noting that the eigenvectors $|n\ra$ form an orthogonal set with respect to the
finite-dimensional analogue of the c-product:
\be
(n|m\ra = \delta_{mn}
\ee
for all $m,n\in \{0,...,N\}$.

The reflection and transmission probabilities are defined as
$|{\cal R}|^{2}:=\lim_{t\to\infty}\sum_{i<\min{\cal I}}|\varphi_{i}(t)|^{2}$ and
$|{\cal T}|^{2}:=\lim_{t\to\infty} \sum_{i>\max{\cal I}}|\varphi_{i}(t)|^{2}$ respectively.
Here, $\varphi_{i}(t)$ is the $i^{th}$ component of $|\Psi(t)\ra$, and $\cal I$ is the region
of non-zero potential.

So far, we have acquired the machinery to
describe the dynamics of a given initial wave function
in the presence of a complex potential by solving
the Schr\"{o}dinger equation.
In the next sections we will utilize it to examine various
physical systems.

\section{Complex potentials as absorbers}

For illustrative purposes, a good example is the complex
Dirac-$\delta$ potential, $V(x) = \lambda\delta(x)$,
where $\lambda\in\compl$ with $\mbox{Im}(\lambda) < 0$ for
an absorbing potential. (The complex square-well potential is also described
in detail in Appendix A.)
Suppose that the one-dimensional space spans from $x=-L$ to $x=L$.
Then, eigenfunctions of the Hamiltonian, $H=p^2/2m+V(x)$, can be classified according to their
parity.
These eigenfunctions must have the form
\be
\psi(x)= \begin{cases} C\cos(kx)+D\sin(kx) & \text{if $-L < x < 0$}
\\ A\cos(kx)+B\sin(kx) &\text{if $0 < x < L$}]\;,
\end{cases}
\ee
where, if $E$ is the corresponding eigenvalue, $k=\sqrt{2mE}$.
For even eigenfunctions, $C=A$ and $D=-B$.
We impose open boundary conditions at $x=\pm L$
(namely, $\psi(-L)= \psi(L) = 0$).
Using the usual matching
conditions for the eigenstates at $x=0$, we find that
\be
\tan(kL) = -\frac{2k}{\lambda}.
\ee
For the odd channel, $C=-A$, $D=B$, and
the eigenstates should vanish at $x=0$, which
requires that $A=0$.
As expected, the odd-channel eigenstates
are independent of $\lambda$. These states are given by
$\psi (x) = \sin(p_nx)$ with $p_n=n\pi/L$.

The initial wave packet,
\be
\Psi_0(x)=(2\pi\alpha)^{-1/4}e^{ik_{0}(x+x_{0})}
e^{\frac{-(x+x_{0})^{2}}{4\alpha}},
\ee
can then be expanded in terms of the eigenfunctions of $H$ by exploiting
their c-product orthogonality:  $\Psi_0=\sum_k c_k \psi_k$, where
\be
c_k = (\psi_k|\Psi_0).
\ee
The time evolution is then given by Eq. (\ref{time evolution}).

For the lattice problem, as described in Sec.~II, one can consider a complex potential
with $U=-i\beta$ at a single site, say $I$. Then the Hamiltonian becomes
$H=-t_{0}\sum_{<i,j>\sigma}C^{+}_{i\sigma}C_{j\sigma}
-i\beta\sum_{\sigma}C^{+}_{I\sigma}C_{I\sigma}$.
Since the Hamiltonian is a complex matrix, the numerical diagonalization
can be done by the driver ZGEEVX contained in the LAPACK
package.\cite{zgeevx}
In Fig.~1, we show the time evolution of a wave packet, which initially
resides at $x_{0} = 50$ with the average momentum $k_{0} = \pi/2$.
The single impurity is at $I = 100$ with $\beta = 1$.
Fig.~2  compares the reflectance and transmittance of the wave packet
with the results from the scattering plane-wave approach for various (average) momenta $k_{0}$.
The scattering plane-wave approach is, in fact, verified by the dynamical calculations. As long as
the wave packet is broad enough, the scattering plane-wave approach
for a complex square well can be verified
in a similar manner.

\begin{figure}[tp]
\begin{center}
\includegraphics[height=3in, width=3in, angle=-90]{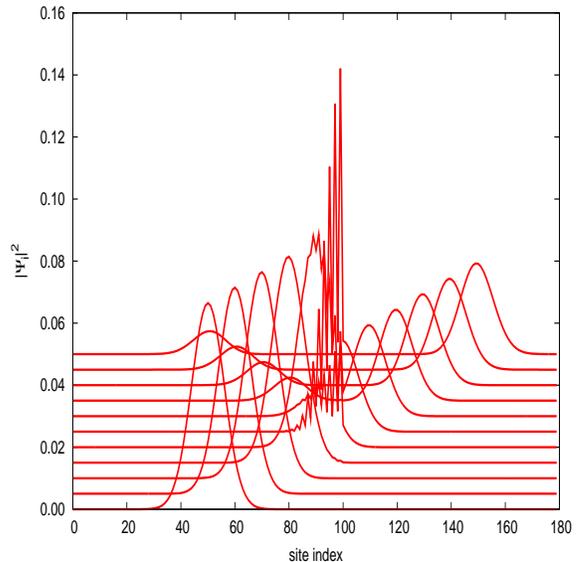}
\caption{(Color online)
The time evolution of a wave packet in the presence of a complex absorbing potential.
}
\end{center}
\end{figure}

\begin{figure}[tp]
\begin{center}
\includegraphics[height=3in, width=3in, angle=-90]{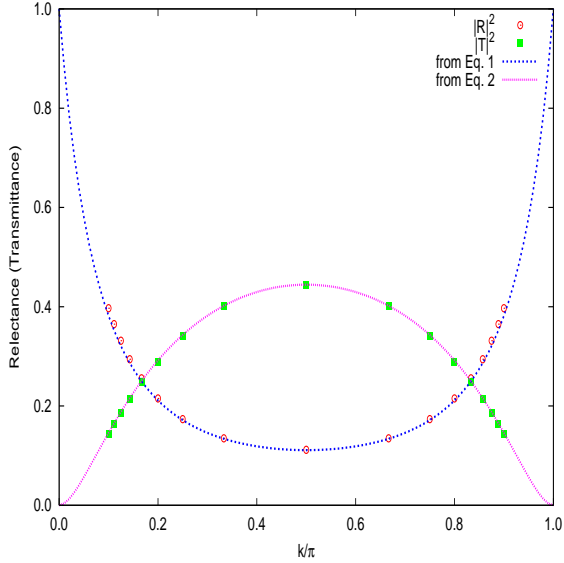}
\caption{(Color online)
The reflectance and transmittance probability as a function of the momentum.
The scattering approach is verified by the full dynamics.
}
\end{center}
\end{figure}

\section{effects of a complex absorbing potential on spin transfer}

We now examine the interaction of an electron with a spin-flip potential followed by a complex
absorbing potential.
In the continuum, a
simple representative Hamiltonian is
$H=p^{2}/2m-2J_{0}{\bf \sigma}\cdot{\bf S}\delta(x)-i\beta\delta(x-a)$,
where $J_{0}$ is the coupling constant of a spin-flip interaction
between the electron spin $\sigma$ and a local spin $S$ at $x=0$.
The distance
between the two interactions, $a$,
is taken to be much larger than the width of an
incoming wave packet. Note the conflicting requirements;
a broad wave packet is required to have a reasonably well-defined momentum,
but a narrow wave packet (on the scale of $a$) is required to observe causal effects
properly. The wave packet
interacts with the local spin first and is partially reflected
and partially transmitted.  (Note that the amplitudes of the transmitted and reflected waves
depend upon the initial spin configuration.)
The transmitted wave later interacts with the complex
absorbing potential after completely emerging from the spin-flip potential. (In other words, the wave
packet does not simultaneously feel both interactions at any given moment.)

It is numerically feasible to find some
eigenvalues for this Hamiltonian; however the comprehensive search for
all complex eigenvalues within some upper modulus bound (given by appropriate truncation limits)
is seen to be a formidable task.  The analogous problem on the lattice is far more tractable, owing
to
the fact that the Hamiltonian is a complex symmetric matrix, which is diagonalizable by
standard numerical routines.
The lattice Hamiltonian is
\be
H=-t_{0}\sum_{<i,j>\sigma}C^{+}_{i\sigma}C_{j\sigma}
-2J_{0}{\bf\sigma}_{l}\cdot{\bf S}_{l}
+\sum_{i\in {\cal I},\sigma}U_{i}C^{+}_{i\sigma}C_{i\sigma}\;,
\ee
where the local spin is at the site $l$ and the complex
absorbing potential sites belong to the set ${\cal I}$ far away
from the site $l$. Since we have only one local spin on the
lattice, we will drop the site label $l$ from now on.

We will investigate the case where initially the local spin is pointing down
$(S_{z}=-S)$ while the electron spin is pointing up
$(\sigma_{z}=1/2)$.  In particular, we wish to monitor
the time evolution of $\la S_{z}\ra$ as the electron interacts with the local
spin and is subsequently partially absorbed by the complex potential.

The expectation value of
a time-independent operator ${\cal A}$ is $\la{\cal A}\ra
=\la\Psi(t)|{\cal A}|\Psi(t)\ra/\la\Psi(t)|\Psi(t)\ra$.
Differentiating with respect to time,
\bwt
\bea
\frac{d}{dt}\la{\cal A}\ra&=&
\frac{d}{dt}\frac{\la\Psi|{\cal A}|\Psi\ra}{\la\Psi|\Psi\ra}
=\frac{\la\Psi|\Psi\ra\frac{d}{dt}\la\Psi|{\cal A}|\Psi\ra
-\la\Psi|{\cal A}|\Psi\ra\frac{d}{dt}\la\Psi|\Psi\ra}{\la \Psi|\Psi\ra^{2}}
\nonumber\\
&=&\frac{i\la\Psi|\Psi\ra\la\Psi|\left[H^{+}{\cal A}
-{\cal A}H\right]|\Psi\ra
-i\la\Psi|{\cal A}|\Psi\ra\la\Psi|\left[H^{+}
-H\right]|\Psi\ra}{\la \Psi|\Psi\ra^{2}}
\nonumber\\
&=&i\la H^{+}{\cal A}-{\cal A}H\ra-i\la {\cal A}\ra\la H^{+}-H\ra
\label{dAdt}
\eea
\ewt
Note that $\la{\cal A}\ra$ is not guaranteed to be constant in time, even if
$[H, {\cal A}] = 0$. If $H$ is Hermitian, this equation reduces
to the usual expression: $d\la{\cal A}\ra/dt=i\la[H, {\cal A}]\ra$.
For example, the $z$-component $J_{z}$ of the total spin
${\bf J}={\bf\sigma}+{\bf S}$ commutes with the Hamiltonian: $[H,J_{z}]=0$.
However, it is clear from Eq.~(\ref{dAdt}) that in general $\frac{d}{dt}\la J_{z}\ra\ne0$.
Consequently, in general, we cannot utilize $\la J_{z}\ra$ as a conserved
quantity. We will further discuss this issue below.

The spin space of this system has a basis of $2(2S+1)$ spin states:
$|\sigma_{z},S_{z}\ra = |+,S\ra,\cdots,|-,-S\ra$. We focus on the simplest case, $S=1/2$,
in which there
are four spin basis states: $|+,\up\ra,\;|+,\dn\ra,\;
|-,\up\ra,\;|-,\dn\ra$. The Hamiltonian matrix is of dimension $(4N\times4N)$, where $N$ is the
number of lattice sites. The initial electron wave packet is also expressed in terms of
these basis states;

\bea
|\Psi(0)\ra =
\sum_j \varphi_{1,j}(0)|+,\up\ra+\varphi_{2,j}(0)|+,\dn\ra& \nonumber\\
+\varphi_{3,j}(0)|-,\up\ra+\varphi_{4,j}(0)|-,\dn\ra&\
\eea

For most of this investigation, only
$\varphi_{2,j}(0)$ is non-zero, with
\be
\varphi_{2,j}(0)=e^{ik_{0}(j-j_{0})}
e^{\frac{-(j-j_{0})^{2}}{4\alpha}},
\ee
 We choose $j_{0}=80$, $k_{0}=\pi/2$, $J_{0}=1$, $\alpha=6$, and
$N=220$. The local spin resides at site $120$, and the absorbing potential sites
are from $180$ to $184$ with a constant $U = -i$ at each site. Fig.~3 shows the
time evolution of both $|\varphi_{2,j}(t)|^{2}$ and $|\varphi_{3,j}(t)|^{2}$.
Part (a) shows the spin-up component of the electron. At the initial scattering
center (local spin at site $120$), a transmitted and small reflected component emerges;
the transmitted component then propagates toward the absorbing potential where some is
reflected, very little is transmitted, and most is absorbed. Part (b) shows the spin-down
component of the electron. There is none until the electron interacts with the local spin.
(Notice the difference in vertical scale compared to (a)). Equal
amounts are propagated to the right and to the left, and eventually most of the packet on
the right is absorbed by the negative imaginary potential.  (As in (a), some is reflected and
some is transmitted as well).

\begin{figure}[tp]
\begin{center}
\includegraphics[height=3in, width=3in, angle=-90]{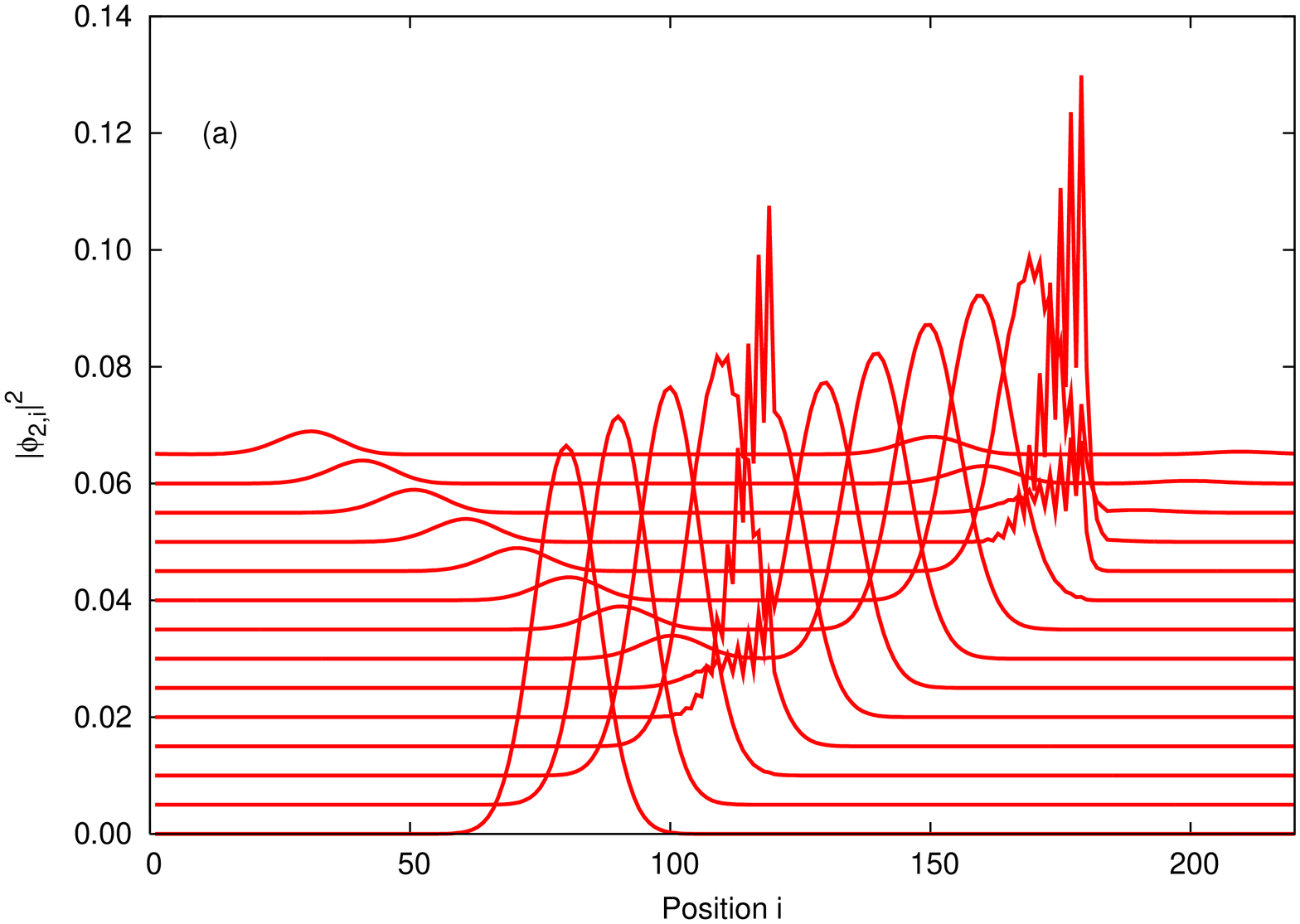}
\includegraphics[height=3in, width=3in, angle=-90]{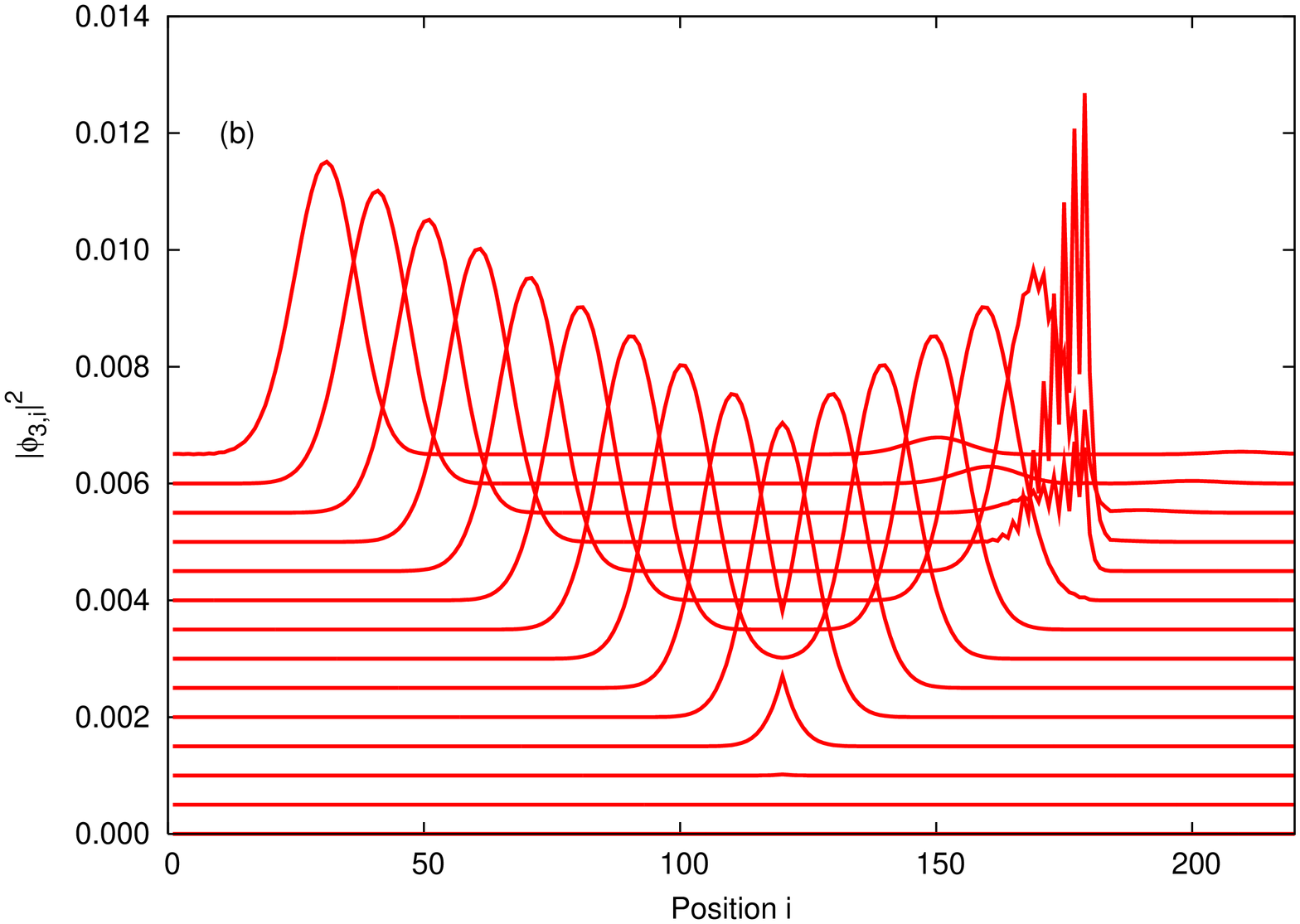}
\caption{(Color online)
The time evolution of the (a) $|\varphi_{2,j}|^2$ and (b) $|\varphi_{3,j}|^2$
components of a wave packet
in the presence of a spin-flip interaction as well as a complex
absorbing potential. The initial mean position of the wave packet
is $x_{0}=80$ with $k_{0}=\pi/2$. The lattice size $N=220$. In this
case, one can actually use
total angular momentum conservation to argue that the other two components are zero.
}
\end{center}
\end{figure}
Each profile in the plot
corresponds to a snapshot of (a) $|\varphi_{2,j}(t)|^{2}$ or (b) $|\varphi_{3,j}(t)|^{2}$,
taken at different times from $t =
0$ to $t = 65$. The wave packet leaves the
local spin at $t = 30$ and does not interact with the complex
potential until $t = 40$. Eventually the transmitted wave packet from
the spin-flip scattering interacts with the complex potentials; by $t=60$ the interaction
is complete. No interaction occurs beyond $t = 60$.

\begin{figure}[tp]
\begin{center}
\includegraphics[height=3in, width=3in, angle=-90]{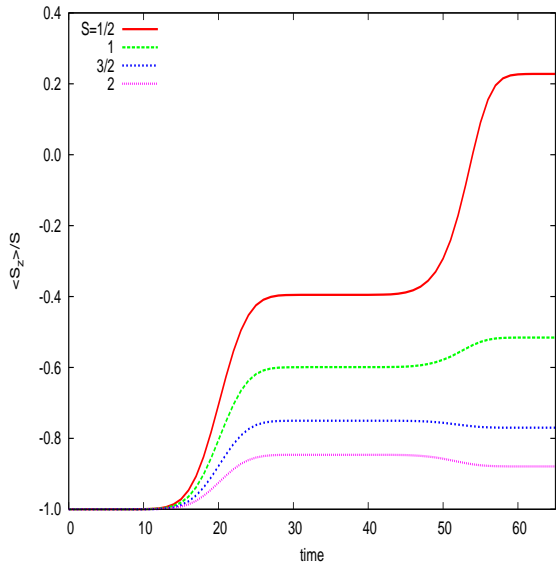}
\caption{(Color online)
The dynamics of $\la S_{z}\ra$ for $S=1/2, 1, 3/2,$ and $2$.
The local spin is at site 120 (reached at approximately time $t=20$).
The spin-1/2 electron is coupled to the local spin with coupling $J_{0}=1$.
The complex potentials are at 5
sites from 180 to 184, reached after time 50; their strengths are a constant, $U=-i$.
}
\end{center}
\end{figure}

The important dynamics of the local spin can also be illustrated by $\la
S_{z}(t)\ra$, which indicates the degree of spin transfer from the
incoming electron to the local spin.\cite{precession} We plot $\la
S_{z}(t)\ra$ as a function of $t$ in Fig.~4 for a variety of values of
S.
In all cases, it is obvious that the
first increase of $\la S_{z}(t)\ra$ is induced by spin transfer from
the incoming electron to the local spin as illustrated well in
Fig.~3b. This is the increase that one will obtain when the
Hamiltonian has only real-valued potentials.
$\la S_{z}(t)\ra$ remains unchanged from $t =
30$ to $t = 40$, which is also consistent with the time evolution of
the wave packet. It is after $t = 40$ that an intriguing feature
takes place when the transmitted wave packet starts interacting with
the complex potentials. $\la S_{z}(t)\ra$ increases considerably even though
(i) the scattering is spin-independent, and, more importantly, (ii) the
wave packet no longer physically overlaps with the local spin.
This phenomenon could be interpreted as a complex potential-driven
action-at-a-distance.
The equation governing $\la
S_{z}(t)\ra$ can be derived based on Eq.~(\ref{dAdt}). Since
$H=H_{0}-iV_{0}$, where $H_{0}=H^{\dagger}_{0}$, and $V_{0}$ is real
valued, one can show
\be
\frac{d}{dt}\la S_{z}(t)\ra=
i\la[H_{0},S_{z}]\ra - 2\left[\la V_{0}S_{z}\ra
-\la V_{0}\ra\la S_{z}\ra\right]\;.
\ee
Since the two interactions do not occur at
the same time because of their arrangement, we can consider the two
separately. The first term gives rise to spin transfer while the
additional increase after t = 40 is attributed to the second term.
As mentioned earlier, if the absorbing potentials were real-valued or zero,
there would be no additional change in $\la S_{z}(t)\ra$. In Fig.~4,
we show $\la S_{z}(t)\ra/S$, where $S=|\la S_{z}(0)\ra|$, for
various values of local spin $S = 1/2$, $1$, $3/2$, and $2$. As
shown in the plot, the effects of the complex potential are most
significant for $S=1/2$.
\begin{figure}[tp]
\begin{center}
\includegraphics[height=3in, width=3in, angle=-90]{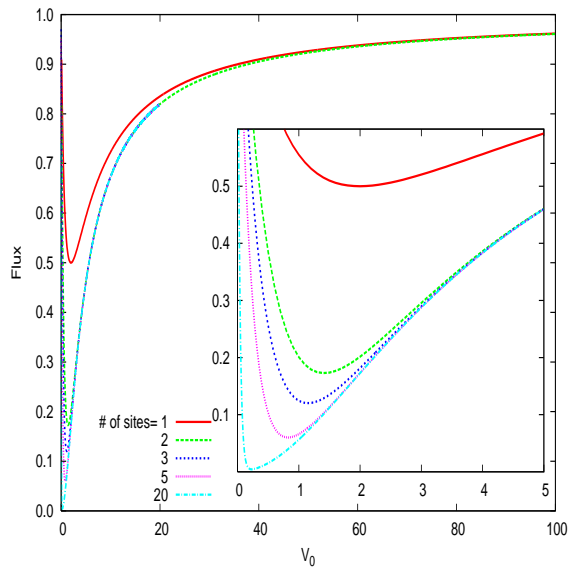}
\caption{(Color online)
Flux reduction for different number of complex potential sites.
}
\end{center}
\end{figure}
On the other hand, for $S\ge3/2$ the
effect of the potential is less considerable and, when the electron interacts
with the absorbing potential, $\la S_{z}(t)\ra$ of the local spin actually {\em decreases}.

Even though the displayed crossover appears for large $S$, a similar crossover can occur
even for $S=1/2$, as will be now explained.
The definition of $\la S_z \ra $ (see above Eq. (\ref{dAdt}))
includes specific combinations of different components of the wave packet.
The relevant components here are those that strike the absorbing potential region; they
remain entangled with the local spin, and hence the spin is affected too.
Let us first examine
what happens when a wave packet strikes an absorbing potential. Fig.~5 illustrates the
flux reduction as a function of impurity potential $U$, for a varying number of impurity
sites (or varying width of imaginary impurity potential well).
Note that we define $U = U_0 - iV_0$, and here we use pure imaginary potentials only.
All imaginary values of the complex potential are taken to be
negative ($V_0 > 0$); positive values result in a flux increase, and are of no interest here.

Analytical expressions can be easily derived for one and two sites, by using the expressions
in the appendix of Ref. \onlinecite{kim1} for $U = -iV_0$. In the case of a single site,
one obtains
\be
{\rm Flux}_1 \equiv  |T|^2 + |R|^2 = 1 - {V_0 \sin{k} \over \sin{k}^2 + V_0 \sin{k}
+ (V_0/2)^2},
\label{single}
\ee
where $k$ is the wave vector of the plane wave.
Note that one obtains unit flux both for $V_0 = 0$ and for $V_0 \rightarrow \infty$; otherwise
there is a sharp reduction as $V_0$ increases from zero, followed by a steady recovery
for increasing values. This expression is plotted (for $k = \pi/2$) in Fig.~6,
and is indistinguishable from the result obtained with a wave packet with finite width ($\alpha =
6$).
The expression for two-impurity sites shows the same characteristics; for $k = \pi/2$ it is
\be
{\rm Flux}_2 =  {4 + V_0^4 \over (2 + 2V_0 + V_0^2)^2}.
\label{double}
\ee
This curve is also indistinguishable from the wave packet result in Fig.~6, and results
for more absorbing potentials differ by very little from the case of two impurities.
Thus it is clear that an optimum imaginary potential strength exists to maximize the flux reduction.
In Fig.~6, in addition, we show the
flux reduction for 5 absorbing impurities, with and without the local spin,
and for two different initial conditions for the local spin. In this plot the flux is normalized to
the
flux transmitted through the local spin.
Finally, in the interest of completeness, one can ask whether a real component of the
absorbing potentials has a strong influence on what has been done so far. In Fig.~7
we show the flux reduction as a function
of the potential $U = U_0 - iV_0$; as before only positive values of $V_0$ are shown. This result is
for a single absorbing impurity. Results for more impurities are similar to this one.
The result for a single impurity is, for $k = \pi/2$,
\be
{\rm Flux} = 1 - {4 V_0 \over (2 + V_0)^2 + U_0^2},
\label{single_complex}
\ee
and agrees perfectly with the corresponding numerical result shown. As Fig.~7
indicates, a large $U_0$ serves to skew the $U_0 = 0$ result. For a single impurity the minimum flux
occurs at $V_{0 {\rm min}} = \sqrt{4 + U_0^2}$ and the flux reduction slowly goes to
zero as $U_0$ increases.

\begin{figure}[tp]
\begin{center}
\includegraphics[height=3in, width=3in, angle=-90]{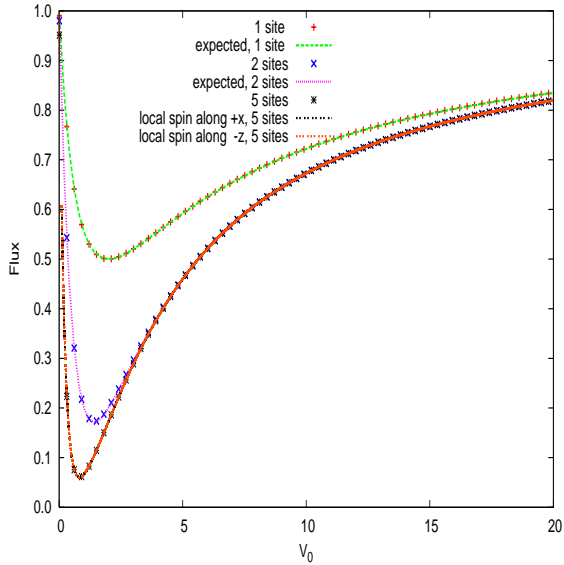}
\caption{(Color online)Flux reduction at the imaginary potential,
with and without local spin. Expected and numerical results for one and two site impurity potentials.
5 sites with local spin with different orientations.} \label{flux_red}
\end{center}
\end{figure}

\begin{figure}[tp]
\begin{center}
\includegraphics[height=3in, width=3in, angle=-90]{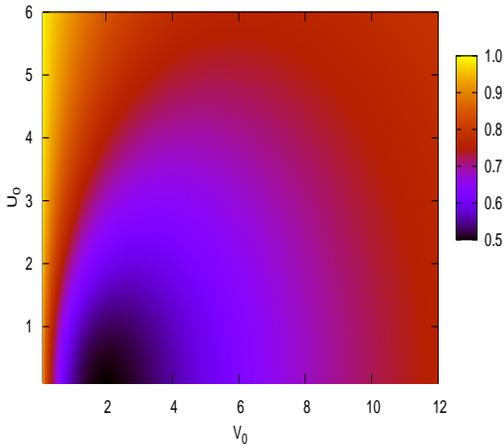}
\caption{(Color online) Flux reduction for complex potential with
both real and imaginary parts.
}
\label{Flux_Real_Imag}
\end{center}
\end{figure}

Returning to the problem that includes the spin-flip interaction,
we can define reflection and transmission
magnitudes that leave the local spin in a particular configuration. For
example, let R1 and R2 be the reflection coefficients with local spin
up and down, respectively,
and let T1 and T2 be the transmission coefficients with local spin up and down
respectively. With these magnitudes the
expectation value of the z-component of the local spin after interaction can be
calculated to be
\be \la
S_z(t)\ra=\frac{R1+T1-R2-T2}{2(R1+T1+R2+T2)}=\frac{1}{2}-\frac{R2+T2}{R1+T1+R2+T2}.
\label{SzTR}
\ee
This equation shows that the final value of $\la S_z \ra$
depends on a particular combination of these magnitudes. Therefore initial
conditions and system parameters, which determine these magnitudes,
change the value of $\la S_z\ra$. An example of this different behavior arises as
a function of the $J_0$ coupling of the electron spin to the local spin. The
expectation value of the z-component of the local spin, $\la S_z \ra$, is shown
as a function of time in Fig.~8, for a variety of values of $J_0$.
As illustrated, the change after $t=50$ alters its characteristic around
$J_0 = 2.3$; namely, instead of an additional increase, $\la S_z \ra$ decreases when
the coupling (to the local spin) is sufficiently strong.
\begin{figure}[tp]
\begin{center}
\includegraphics[height=3in, width=3in, angle=-90]{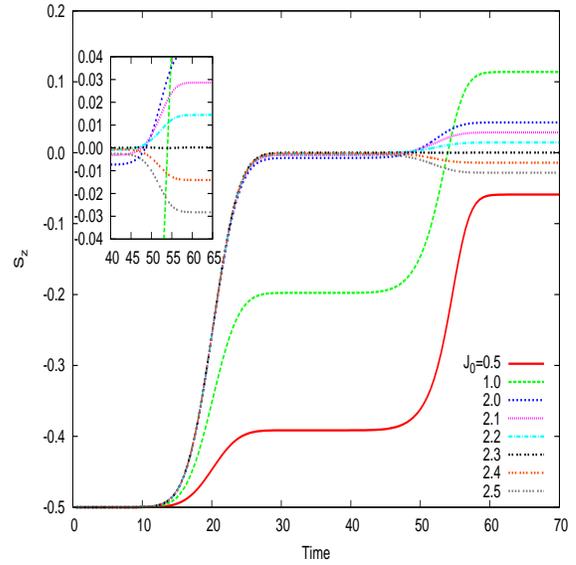}
\caption{(Color online)
Time evolution of $\la S_z\ra $ of local spin with an electron spin up interacting with local spin
and imaginary potential with different electron-spin interaction strength.
}
\end{center}
\end{figure}

As was the case with varying magnitude of spin (see Fig. 4), the local spin can
react in a qualitatively very different way when the transmitted part of the
electron reaches the absorbing potential.
A summary of the scattering coefficients after interaction with the local spin is displayed in
Fig.~9.
Note that $J_0 \approx 2.3$ plays a prominent role; it is the value of coupling
for which the spin-flip component of the transmitted wave packet peaks as a function of
$J_0$. One can readily show that the change in $\la S_z(t)\ra$ as given by Eq. (\ref{SzTR})
after scattering from an imaginary potential (compared with the change before scattering) is
proportional
to $T2 - R2$. This quantity changes sign at $J_0 \approx 2.3$, and therefore leads to
the qualitative change indicated in Fig.~8, independent of the value of the absorbing potential.
\begin{figure}[tp]
\begin{center}
\includegraphics[height=3in, width=3in, angle=-90]{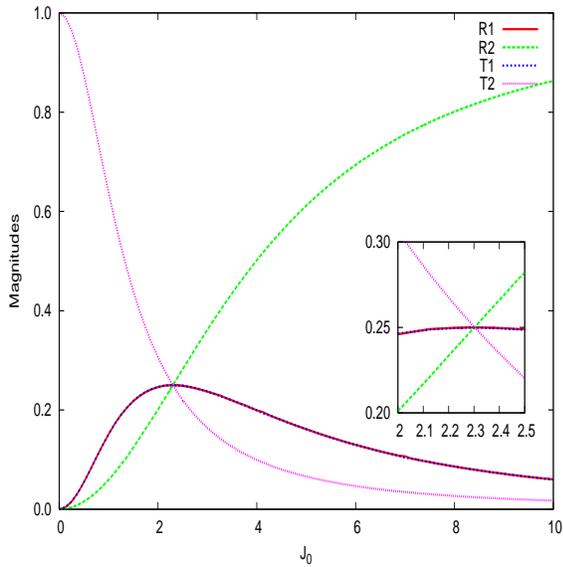}
\caption{(Color online) Magnitudes of each component of the
wave packet with respect to local spin. R1 has the same value as
T1, as the created state of electron down, local spin up needs to conserve momentum.}
\label{Ref_Tra}
\end{center}
\end{figure}

\begin{figure}[tp]
\begin{center}
\includegraphics[height=3in, width=3in, angle=-90]{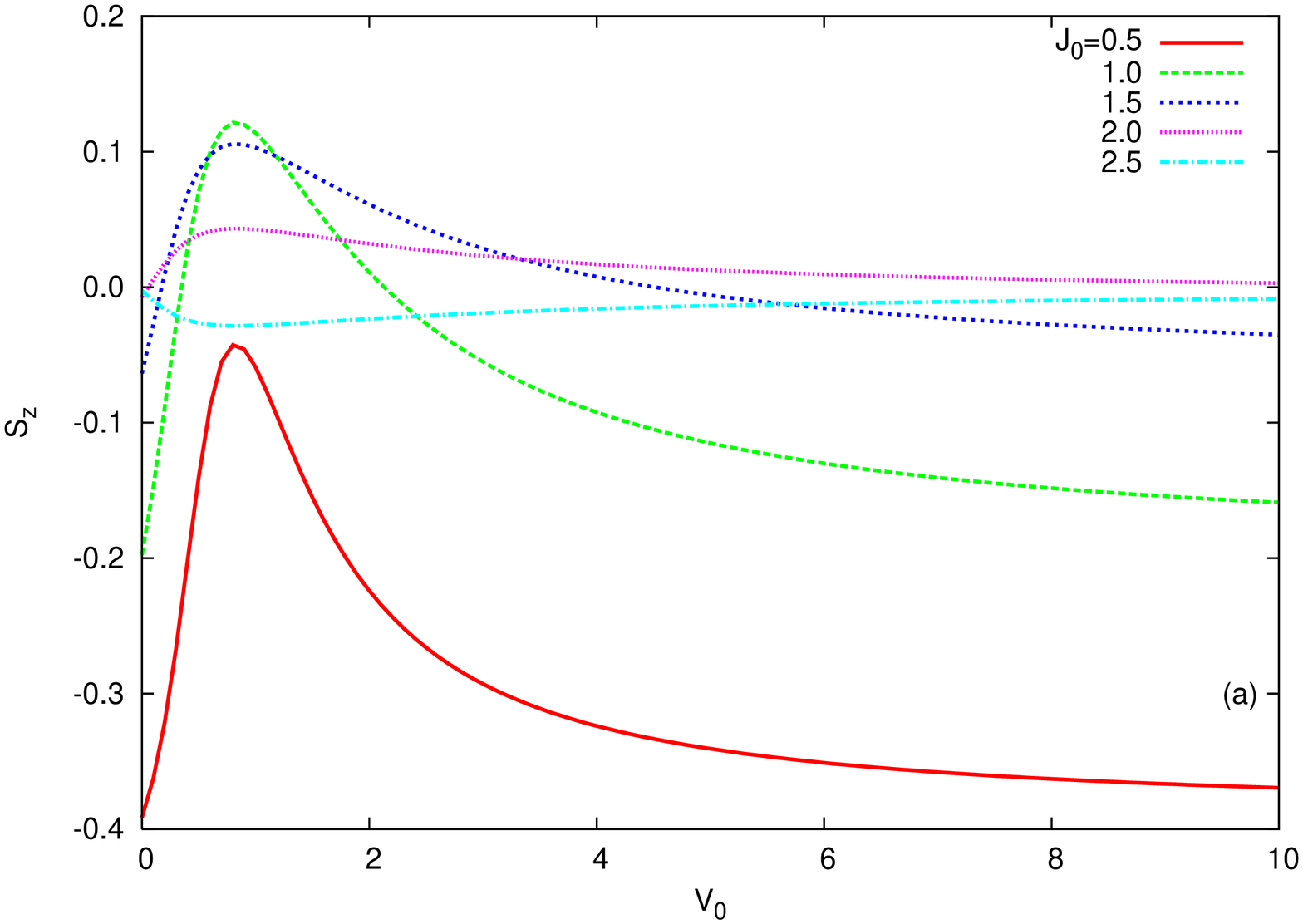}
\includegraphics[height=3in, width=3in, angle=-90]{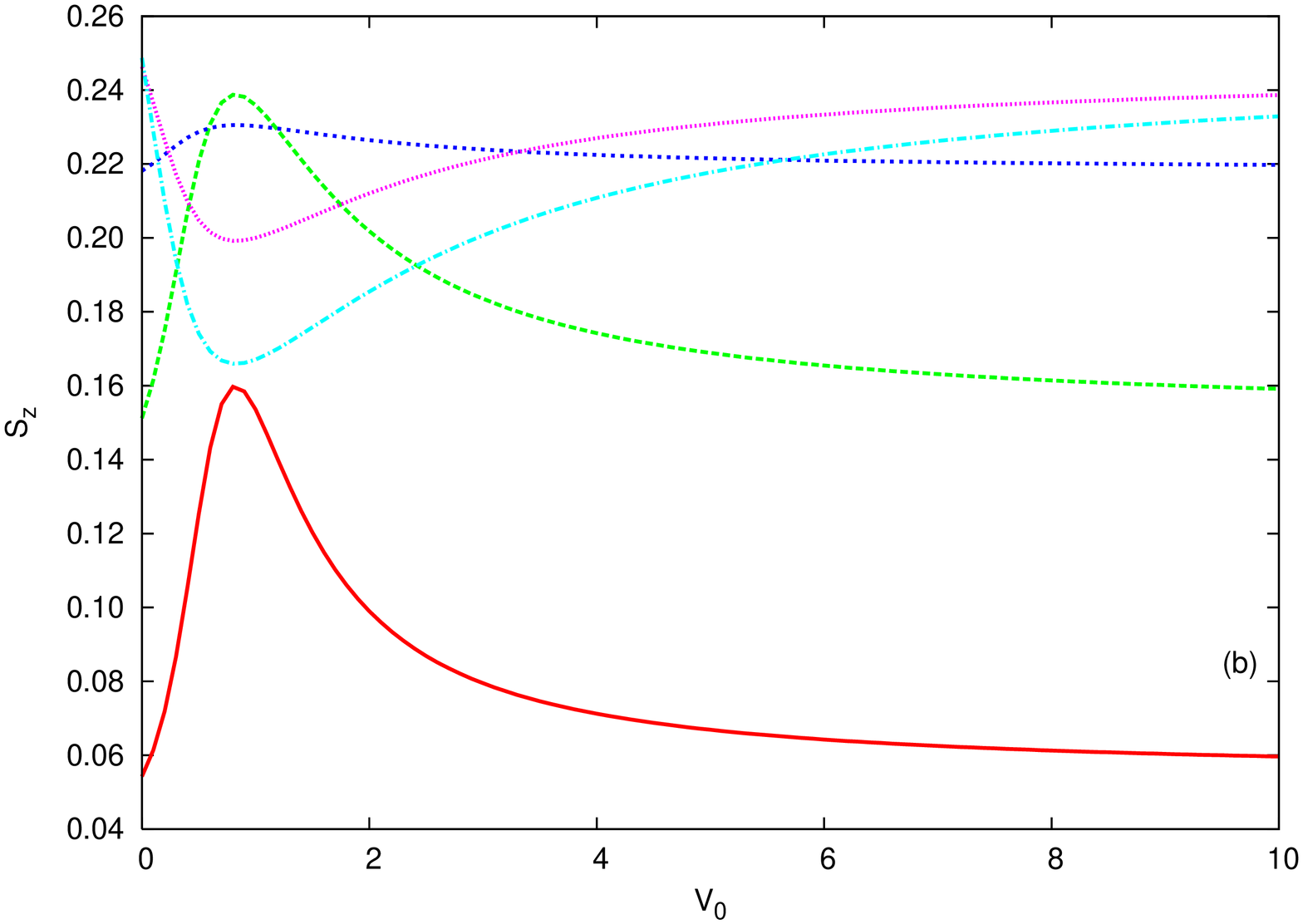}
\caption{(Color online) z-component of local spin for different
interaction strengths of electron-spin and imaginary potential, with
initial configuration of (a) $|+\dn\ra$, and (b) $|+0\ra$. }
\label{Sz_z}
\end{center}
\end{figure}

The impact of the magnitude of $V_0$ (negative imaginary absorbing potential) on the magnitude of
change
is illustrated in Fig. \ref{Sz_z}, for a couple of local spin initial configurations, (a) $\la S
\ra$
initially in the -z-direction, and (b) $\la S \ra$ initially in the x-direction.

In figure \ref{Sz_z}(a),  $\la S_z \ra$ is plotted for a
time after the transmitted part of the wave packet interacts with the
imaginary potential, for the case where the initial local spin configuration
was aligned in the negative z-direction (the incoming electron is always spin up).
For low values of $J_0$, the imaginary potential first causes an increase in the
local z-component spin, followed by a decrease (for large enough $V_0$ the local
spin is not affected, which is consistent with Fig. 5). As already noted, for larger
$J_0$ (greater than about $J_0 \approx 2.3$ in the figure), the effect of the absorbing
potential is opposite: the local z-component spin first decreases and then increases
as a function of $V_0$.

\begin{figure}[tp]
\begin{center}
\includegraphics[height=3in, width=3in, angle=-90]{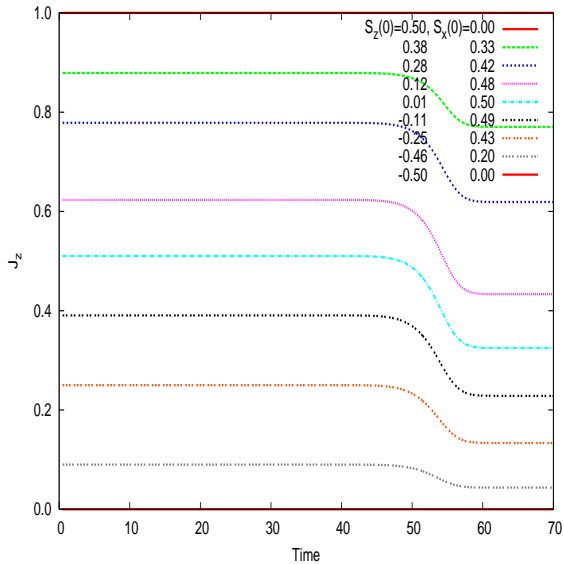}
\caption{(Color online) $\la J_z\ra$ of the local spin for various local spin initial conditions
as a function of time. $J_0 = 1$ and $V_0 = 1$.}
\label{Jz_time}
\end{center}
\end{figure}

Note that the initial configuration described above is, in many ways, quite unique, because
the local spin and the electron spin are perfectly anti-aligned initially. It turns out
that for this initial configuration (and for some others), $J_z$ is conserved, contrary
to the general expectation based on Eq. (\ref{dAdt}) (with ${\cal{A}} = J_z$).
In Fig.~\ref{Sz_z}(b) we show an example where $J_z$ is not conserved, as one would expect in
general.
To show this more explicitly, in Fig.~11 we plot a sequence of results for initial starting
configurations that sweep through the x-z plane.
At the two end-points $J_z$ remains constant as a function of time.
For other initial conditions, $\la J_z\ra$ changes as a function of time.
As Fig. \ref{flux_red} showed, the amount of flux reduction due to the absorption
is dependent on the value of $V_0$. For large $V_0\rightarrow\infty$,
the reduction is diminished by the reflection from the potential,
and full flux is recovered. This effect is visible in $\la S_z\ra$ as well. The
main change occurs around $V_0\approx 1$, and full recovery (to the $V_0=0$ value)
occurs for high values of the interaction.

\section{summary}
Complex absorbing potentials have been utilized as a source of absorption in the
context of dynamical problems (like spin-flip scattering) by investigating the time
evolution of a particle wave packet under the influence of such potentials.
The formulation presented in this paper is mathematically equivalent
to diagonalizing a complex symmetric matrix because the Hamiltonian
we consider is, in the matrix representation, complex symmetric
rather than Hermitian. The structure of the formalism parallels the one in 'conventional'
quantum mechanics with Hermitian Hamiltonians. However, the
diagonalizability of complex symmetric matrices requires some extra attention.
Eq. (\ref{dAdt}) is central to the results when a complex potential is considered.
This equation indicates that conservation of a physical quantity does not follow from
the commutability of its operator with the Hamiltonian in the presence of a complex
(i.e. non-Hermitian) potential. One of the most
remarkable examples is the total angular momentum ${\bf J}$.
We showed an example where $\la J_{z}\ra$ is not conserved, and shows significant
time dependence, depending on the initial conditions.

We also studied the interplay between the spin-flip potential and the absorbing potential.
The first, while Hermitian, nonetheless displays some unusual properties connected to
entanglement. Essentially, the electron wave packet that emerges from the spin flip scattering
cannot be separated from the local spin state. As a consequence,
when the transmitted portion further interacts with an absorbing potential,
the state (and expectation values) of the local spin is affected. We also studied the
'absorbing power' of a complex potential as a function of its strength (both real and imaginary
parts).
The remarkable 'action-at-a-distance' displayed in Figs. 4 and 5 is a consequence
of an imaginary potential. It would be most intriguing if this effect can
be observed in a real experiment.

A complex absorbing potential for matter waves can be realized by a standing
light wave resonant with a transition between a metastable state and an excited
state, which corresponds to an open two-level system. Such a scheme has been
used in various experiments\cite{oberthaler,bernet,stutzle} in quantum optics.

Theoretically, complex potentials are used frequently in quantum mechanical
problems to simulate absorption processes. Nonetheless all of their effects including the one we
illustrate here may not be completely  physical. However, this is far from clear and requires an
experimental test to delineate the possibilities. As mentioned above, the most promising
candidate for a physical realization is a quantum optical system.

\begin{acknowledgments}

We would like to thank Lucian Covaci for discussions on various aspects
of this work. This work was supported in part by
the Natural Sciences and Engineering Research Council of Canada (NSERC),
NSERC, by ICORE (Alberta), by the Canadian Institute for Advanced Research
(CIfAR), and by the Robert Welch Foundation (grant no. E-1146). FM is grateful to the Aspen Center
for Physics, where some of this work was done.

\end{acknowledgments}


\appendix

\section{complex square-well potential}

For a complex square-well potential, $V(x) = V_{0}$ for $|x|<a$
while $V(x)=0$ elsewhere.
The general solution to the eigenvalue equation,
$H\psi = E\psi$, can be expressed as follows:
\be
\psi(x)= \begin{cases} C\cos(kx)+D\sin(kx) & \text{if $x<-a$,}
\\ F\cos(\kappa x)+G\sin(\kappa x) & \text{if $-a<x<a$,}
\\ A\cos(kx)+B\sin(kx) &\text{if $x>a$,}
\end{cases}
\label{eigenstate_in_Appndx}
\ee
where $A,B,C,D,F$ and $G$ are complex constants,
$\kappa=\sqrt{2m(E-V_{0})}$ and $k=\sqrt{2mE}$.
As the Hamiltonian is non-Hermitian, both $k$ and $\kappa$
lie in the complex plane.
Since the potential $V(x)$ is even, the Hamiltonian operator commutes with
the parity operator.  Hence, all non-degenerate eigenstates of the Hamiltonian
must also be eigenstates of parity.

Consider first the even states.
Evenness of $\psi$ implies that $A=C$, $D=-B$, and $G=0$ in
Eq.~(\ref{eigenstate_in_Appndx}).
The derivative $\partial_{x}\psi$ must be odd,
so $\partial_{x}\psi|_{L}=-\partial_{x}\psi|_{-L}$.
However, periodic boundary conditions require that
$\partial_{x}\psi|_{L}=\partial_{x}\psi|_{-L}$.  Therefore,
\be
\partial_{x}\psi|_{L}=k\left(-A\sin(kL)+B\cos(kL)\right)=0.
\ee
By the continuity of $\psi$ and $\psi\prime$ at $x=a$,
\be
F\cos(\kappa a)+G\sin(\kappa a) = A \cos(k a) + B \sin(k a),
\ee
\begin{center}
and
\end{center}
\be
\kappa \left(-F\sin(\kappa a) + G\cos(\kappa a)\right) =
k \left( -A\sin(ka)+B\cos(ka)\right)\;.
\ee
It is easy to search for eigenvalues which satisfy $\cos(kL)=0$,
$\cos(\kappa a)=0$, or $\cos(ka)=0$, since the zeros of these functions
are well-known.
However, in most configurations, the vast majority of eigenvalues
do not satisfy these conditions, so Eq. (20) must be solved numerically
in the complex plane.  This is done by applying Newton's method
to the following function:
\be
f_{n}(k)= \text{tan} ^{-1}
\left(\frac{\kappa}{k}\tan(\kappa a)\right)+k(L-a)-n\pi\;,
\ee
for each $n$.

Analysis of the odd states is similar to that of the even states.
Oddness implies that $D=B$, $C=-A$ and $F=0$ in
Eq.~(\ref{eigenstate_in_Appndx}).
We apply periodic boundary conditions and require continuity
of $\psi$ and $\psi\prime$ at $x=a$ in order to derive the
following expression for odd eigenstate momenta $k$.
If $\cos(kL)\neq 0$, $\cos(\kappa a) \neq 0$, and $\cos(ka)\neq 0$,
we obtain
\be
k \tan(\kappa a)=-\kappa \tan k(L-a)\;.
\ee

In order to solve this equation in the complex plane,
Newton's method is applied to the function
\be
g_{n}(k) = \text{tan}^{-1}\left(\frac{k}{\kappa}
\tan(\kappa a)\right) + k(L-a) - n \pi\;,
\ee
for each $n$.
It can be seen numerically that in most cases,
$f_{n}(k)$ and $g_{n}(k)$ each have one zero per value of $n$.
Occasionally, one can expect to find 0 or 2 zeros for some $n$.

Let us consider the time-evolution of a Gaussian wave packet
with average position $-x_{0}$ and average momentum $k_{0}$:
\be
\Psi(x,0)=(2\pi\alpha)^{-1/4}e^{ik_{0}(x+x_{0})}
e^{\frac{-(x+x_{0})^{2}}{4\alpha}}.
\ee
Here, $(x_{0}-a)/\alpha$ is taken to be
sufficiently large so that $\Psi(x,0)\approx0$ if $x>-a$.
Propagation of this wave packet under the influence of
the potential $V(x)$ cannot be described semi-classically
due to the discontinuities of $V$.
However, modeling is possible using the method described so far.
The wave packet is expanded in terms of eigenstates as described in Sec.~II
with coefficients as follows:
\bwt
\be
C_{n}= \pm \frac{1}{\sqrt{\mathcal{N}}}
\left ( \frac{\pi\alpha}{2} \right) ^{1/4}
        \left[ (A+iB)e^{-ikx_{0}}e^{-\alpha (k_{0}+k)^{2}}
      + (A-iB)e^{ikx_{0}}e^{\alpha (k_{0}-k)^{2}} \right]\;,
\ee
\ewt
where the $\pm$ in front implies '$+$' for even states
and '$-$' for odd states, and $\mathcal{N}=(n|n)$
is a normalization factor. For each configuration,
we manually check that $(n|n) \neq 0$ for the state involved
in the expansion.

\bibliographystyle{prb}

\begin{thebibliography}{1}

\bibitem{muga} J.G. Muga, J.P. Palao, B. Navarro, I.L. Egusquiza,
Phys. Report {\bf 395} 357 (2004).

\bibitem{horn} R.A. Horn and C.R. Johnson,
{\it Matrix Analysis} (New York, Cambridge University Press, 1985).

\bibitem{vibok} A. Vibok and G. Balint-Kurt, J. Chem. Phys. {\bf 96},
7615 (1992).

\bibitem{bender} C.M. Bender and S. Boettcher, Phys. Rev. Lett. {\bf 80},
5243 (1998).

\bibitem{midgley} S. Midgley and J.B. Wang, Phys. Rev. E {\bf 61} 920 (2000).

\bibitem{rasmussen} A.J. Rasmussen, S.J. Jeffrey, and S.C. Smith,
Chem. Phys. Lett. {\bf 336} 149 (2001).

\bibitem{neumair} A. Neumaier and V.A. Mandelshtam, Phys. Rev. Lett. {\bf 86}, 5031 (2001).

\bibitem{santra} R. Santra and L.S. Cederbaum, Phys. Reports {\bf 368}, 1 (2002).

\bibitem{moiseyev} N. Moiseyev, S. Scheit, and L.S. Cederbaum,
J. Chem. Phys. {\bf 121}, 722 (2004).

\bibitem{moiseyev2} N. Moiseyev, Phys. Reports {\bf 302}, 212 (1998).

\bibitem{riss93} U.V. Riss and H.-D. Meyer, J. Phys. B: At. Mol. Opt. Phys. {\bf 26} 4503 (1993).

\bibitem{riss96} U.V. Riss and H.-D. Meyer, J. Chem. Phys. {\bf 105} 1409 (1996).

\bibitem{kim1} W. Kim, L. Covaci, and F. Marsiglio,
Phys. Rev. B {\bf 74}, 205120 (2006).

\bibitem{postulation} As in quantum mechanics with a Hermitian
Hamiltonian, we use the expansion postulate in this paper.
For the formulation of this postulate see any text book on
Quantum Mechanics, such as {\em Principles of Quantum Mechanics},
by R. Shankar (Plenum Press, New York, 1994, 2nd edition).

\bibitem{self_orthogonal} Note that we assume that
there are no self-orthogonal states, which satisfy $(n|n)=0$.
In that case, our analysis would not necessarily be applicable.
We have checked that, for the physical systems considered in this
paper, there are no such states.

\bibitem{truncate} It is practically required to truncate
the expansion of the wave packet
in terms of eigenstates for some cases in the continuum limit
because the Hilbert space is infinite.

\bibitem{zgeevx} Some modification is needed
to make the eigenstates satisfy the $c$-product
normalization because the numerical driver provides the eigenstates with
the Euclidean normalization. This procedure is, in fact, helpful
to make sure that no eigenstates are self-orthogonal.

\bibitem{precession} In order to see the precession of the local spin,
one needs to evaluate $\la S_{+}(t)\ra$, where $S_{+}=S_{x}+iS_{y}$.

\bibitem{oberthaler} M.K. Oberthaler, R. Abfalterer, S. Bernet, C. Keller,
J. Schmiedmayer, and A. Zeilinger, Phys. Rev. A {\bf 60}, 456 (1999).

\bibitem{bernet} S. Bernet, R. Abfalterer, C. Keller, M.K. Oberthaler,
J. Schmiedmayer, and A. Zeilinger, Phys. Rev. A {\bf 62}, 023606 (2000).

\bibitem{stutzle} R. St{\"u}tzle, M.C. G{\"o}bel, Th. H{\"o}rner, E. Kierig,
I. Mourachko, M.K. Oberthaler, M.A. Efremov, M.V. Fedorov, V.P. Yakovlev,
K.A.H. van Leeuwen, and W.P. Schleich, Phys. Rev. Lett. {\bf 95}, 110405
(2005).

\end{thebibliography}

\end{document}